      [1999/12/01 v1.4c Il Nuovo Cimento]
\documentclass{cimento}
\usepackage{graphicx}
\newcommand{\pb}{pb$^{-1}$}
\newcommand{\fb}{fb$^{-1}$}
\newcommand{\sqrts}{$\sqrt{s}$}
\newcommand{\ttbar}{$t\bar{t}$}

\newcommand{\pt}{$p_T$}

\title{Top quark physics prospects with 10~$fb^{-1}$ at 7~TeV}
\author{A.~Giammanco\from{ins:ucl}\from{ins:fnrs}\from{ins:nicpb}}
\instlist{
	\inst{ins:ucl} Universit\'e Catholique de Louvain, Louvain-la-Neuve, Belgium
  	\inst{ins:fnrs} Fonds National de la Recherche Scientifique (FNRS), Belgium
	\inst{ins:nicpb} National Institute of Chemical Physics and Biophysics, Tallinn, Estonia
}
\PACSes{\PACSit{14.65.Ha}{Top quarks}}
\begin{document}

\maketitle

\begin{abstract}
This paper presents a personal point of view on the prospects open for top quark physics analyses at the LHC with 10~fb$^{-1}$ of pp collisions at 7~TeV. We discuss the main directions for improvement for the analyses presented at TOP2011 when moving to a regime where systematic uncertainties are the limiting factors, and suggest some new studies that become possible for the first time thanks to the larger statistics.
\end{abstract}

\section{Introduction}

The top quark is the heaviest elementary particle discovered so far, and in many ways it is a very uncommon quark. The fact that its electroweak decay is faster than the hadronization timescale implies that the top quark exists only as a free quark, so that the effects from new physics could show up very clearly by comparing measurements with the precise Standard Model preditions.
Its ``re-discovery'' at the Large Hadron Collider (LHC) has been a major milestone for the ATLAS~\cite{atlas} and CMS~\cite{cms} experiments, due to the complexity of its final state which demands a good control of the experimental apparatus. Its expected large coupling to the SM Higgs boson will also be relevant as a test of the Standard Model.

ATLAS and CMS recorded more than 5 \fb\ of pp collisions delivered by the LHC at 7~TeV during 2011.
At the time of writing, the plans for running in 2012 are not final and the beam energy itself is still under discussion. Here we assume that at least 10 \fb\ will have been collected at 7 TeV by the time of the next TOP workshop, and we discuss the kind of analyses that become possible with this amount of data.

\section{Top quark pair production}
\label{sec:tt}

\begin{figure}
\includegraphics[width=0.40\columnwidth,angle=90]{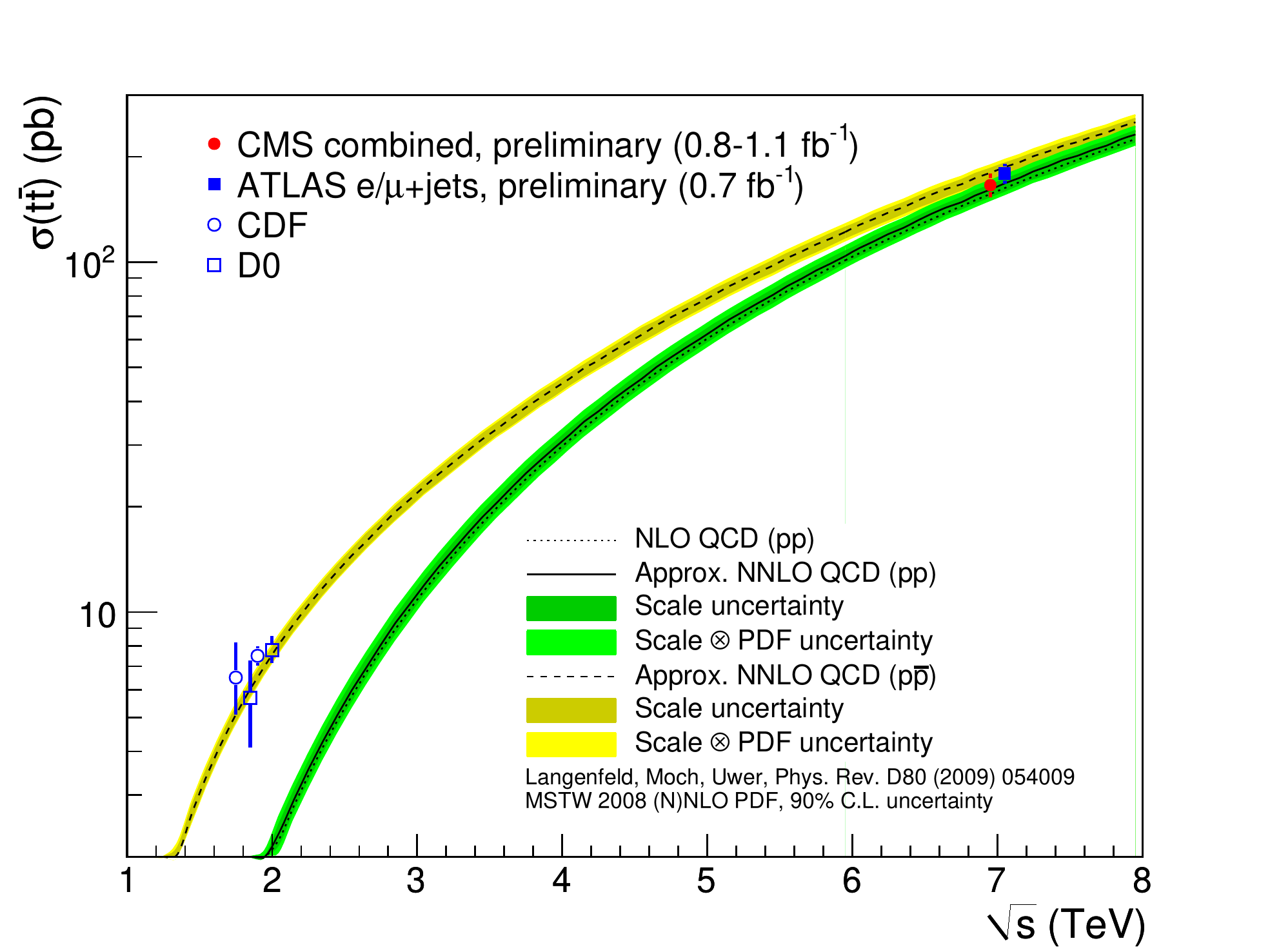} 
\includegraphics[width=0.55\columnwidth]{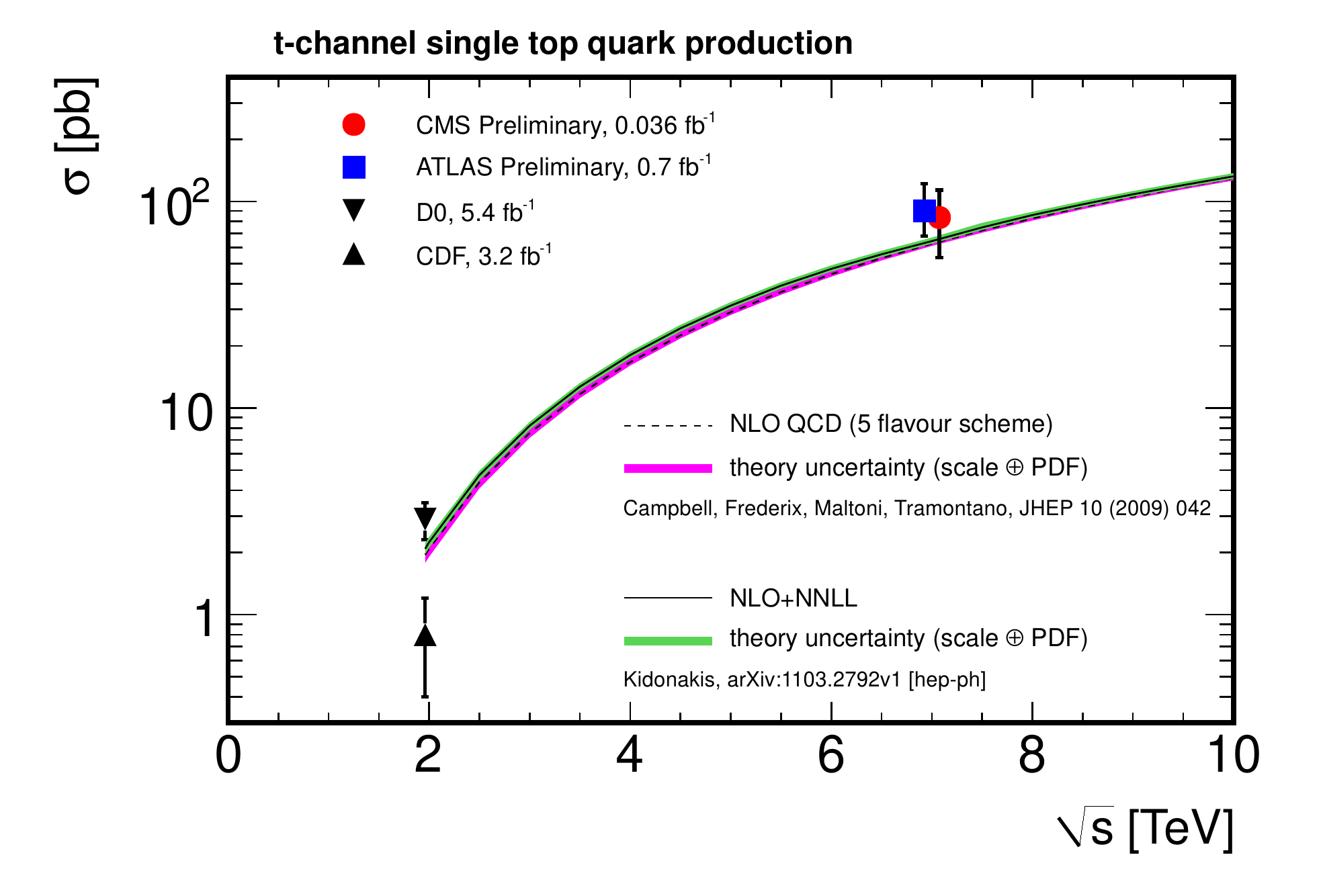} 
\caption{\label{fig:sqrts}Cross section measurements  at Tevatron and LHC and expectations as a function of $\sqrt{s}$ for top quark pair production (left) and t-channel single top quark production (right).}
\end{figure}

The inclusive cross section for pair-produced top quarks (\ttbar) is now predicted at Next-to-Leading Order (NLO) in perturbative QCD, and several approaches have been pursued to approximate the Next-to-Next-to-Leading Order (NNLO)~\cite{uwer}. The theoretical uncertainty for the cross section of pair-produced top quarks in pp collisions at 7 TeV is around 10\%, dominated by the uncertainty on the parton distribution functions (PDF).
The corresponding measurements performed by ATLAS~\cite{xsec_ttbar_atlas} and CMS~\cite{xsec_ttbar_cms} (figure~\ref{fig:sqrts}, left) already reached a similar precision with data sets of around 1 \fb, starting to challenge theory.
 Although the statistical uncertainty is not anymore one of the leading uncertainties of these measurements, larger data sets will indirectly yield benefits: by allowing to trade off statistics for purity with tighter selection thresholds, and by reducing uncertainties coming from the sidebands used to constrain the main systematic uncertainties.
 The uncertainty on luminosity of about 5\% is very unlikely to improve significantly; the solution would be to quote ratios of the \ttbar\ cross section over those of W and Z boson production, whose comparison with theory represents a good handle to constrain PDFs.
 The gluon PDF can be constrained very effectively by measuring differential cross sections versus the rapidity and \pt\ of the \ttbar\ system.

A differential distribution of great importance is the $M_{t\bar t}$ spectrum. 
 Mostly used so far by both collaborations to set limits on narrow resonances that decay to \ttbar~\cite{mtt_atlas1,mtt_atlas2,mtt_cms1,mtt_cms2}, this distribution might be deformed in many ways by new physics (e.g., tail enhancement, peak-dip from interference, etc.); assuming the SM, it can be used for an alternative measurement of the mass of the top quark~\cite{frederix2009}.
 The boosted-top tagging technique~\cite{boosted1,boosted2} applied by CMS in the all-hadronic channel~\cite{mtt_cms2} could be extended to the hadronic side in the lepton+jets final state, further enhancing the sensitivity of this search by the addition of a smaller but independent data sample.

Recently, the CDF and D0 experiments reported tantalizing hints of a significant forward-backward asymmetry between top and anti-top quarks in \ttbar\ production with $p\bar p$ collisions at 1.96~TeV~\cite{afb_cdf1,afb_cdf2,afb_d0}, larger than the expectation from NLO QCD, stimulating a large production of theoretical papers to account for the excess in terms of new physics~\cite{aguilar}.

 No forward-backward asymmetry is observable at LHC because of the charge symmetry of the initial state, but both CMS~\cite{asym_cms} and ATLAS~\cite{asym_atlas} exploit the difference in average momentum between quarks and antiquarks in the initial state in $pp$ collisions, and report differential distributions in $\Delta y = |y_t - y_{\bar t}|$ (ATLAS), $\Delta y^2 = y_t^2 - y_{\bar t}^2$ and $\Delta \eta = |\eta_t - \eta_{\bar t}|$ (CMS). These distributions are unfolded to correct for selection and reconstruction inefficiencies and biases and therefore allowing an easier comparison with theory predictions from new physics models.

 Both experiments also report integrated asymmetries in these variables, i.e.,
\begin{eqnarray}
A_C^y &=& \frac{N(\Delta y > 0)-N(\Delta y<0)}{N(\Delta y > 0)+N(\Delta y<0)},\\
A_C^\eta &=& \frac{N(\Delta \eta > 0)-N(\Delta \eta<0)}{N(\Delta \eta > 0)+N(\Delta \eta<0)},
\end{eqnarray}
 finding values compatible with zero and with the small SM expectation. CMS also reported $A_C^y$ and $A_C^\eta$ versus $M_{t\bar t}$, finding no hint for any dependence.
 The greatest challenge for analyses at the LHC comes from the fact that the majority of partonic collisions at 7 TeV involve two gluons, while a charge asymmetry is typically expected to come from $q\bar q$ initial states.
 With more statistics, LHC analyses will be able to correlate the asymmetry measurement with the $q\bar q$ fraction and with initial state radiation, for example by measuring $A_C^y$ and $A_C^\eta$ as a function of the \pt\ or boost of the \ttbar\ system~\cite{aguilar-juste-rubbo}, and the number of associated jets.

Associated production ($t\bar t X$) will provide important constraints on the couplings of the top quark. With 10 \fb, the couplings to heavy vector bosons will be measurable thanks to a sizeable number of clean $t\bar t Z$ and $t\bar t W$ events. The ATLAS collaboration presented a first study of $t\bar t \gamma$ events~\cite{ttgamma_atlas} for photon energies above 8 GeV, yielding a result compatible with the SM expectation and incompatible with the no-photon hypothesis at $2.7\sigma$ level. With more statistics, in addition to providing a measurement of the charge of a naked quark, this channel will yield constraints to excited top quarks ($t^*\to t\gamma$).

 The most interesting associated production is arguably the process $t\bar t H$, which is sensitive to the Yukawa coupling of the Higgs boson to the top quark and therefore would test the SM correspondence between $Y_t$ and $m_t$, a particularly important issue given the strikingly high value of the latter and the suggestive ``coincidence'' of $Y_t\approx 1$ as inferred from the top quark mass.
 The main challenges of the $t\bar t H$ search are the large combinatorics (especially in the $H\to b\bar b$ decay channel, favoured at low Higgs boson masses) and the difficulty to model the $t\bar t b\bar b$ and $t\bar t + N~jets$ backgrounds; comparatively little effort has been devoted by the LHC collaborations to this search, after simulation studies of the prospects with realistic conditions concluded that very large data sets would have been necessary for discovery in this channel~\cite{tdr_cms,tdr_atlas}.
 More recently, though, the improvement of several analysis tools, which already allowed ATLAS and CMS to outperform their own expectations in a wide range of physics topics, and the development of new ones, like the algorithms for boosted objects~\cite{boosted2}, induce to think that sensitivity to a light Higgs boson associated to \ttbar\ might come much earlier than foreseen by these studies~\cite{ttH}.
 A guaranteed outcome, in any case, will be a better understanding of the $t\bar t b\bar b$ and $t\bar t + N~jets$ processes (a first ATLAS study has been reported for the latter~\cite{ttNj_atlas}), which constitute by themselves a precious testbed for perturbative QCD.

\section{Single top quark production}
\label{sec:st}

Single top quark production, first observed by CDF and D0 in 2009~\cite{gerber}, has been quickly confirmed by CMS and ATLAS in t-channel production~\cite{st_cms,st_atlas} (figure~\ref{fig:sqrts}, right).
 Tevatron and LHC collisions give access to different mixtures of electroweak single-top processes~\cite{motylinski}: t-channel and s-channel production contribute less evenly at 7 TeV (65 pb and 5 pb respectively) than at 1.96 TeV (2 pb and 1 pb), and $W$-associated production ($tW$) which is virtually inaccessible at the Tevatron becomes the second-largest contributor at the LHC (15 pb), and a significant background in \ttbar\ studies and Higgs boson searches. Both collaborations have presented their first results on $tW$~\cite{tw_cms,tw_atlas} and ATLAS has set an upper limit (roughly five times the SM prediction) on the s-channel process~\cite{ss_atlas}.
 All these measurements are already limited by systematic uncertainties, and most of the planned developments will focus on strategies for an {\it in situ} control of the main systematics, and on better background rejection, mostly by means of multivariate analyses.

Precision measurements of all three production modes will have a deep impact  on PDF constraints, with the three channels being complementary to each other and also to \ttbar\ production. For example, t-channel and $tW$ cross sections are sensitive to the $b$-quark PDF and anticorrelated with the $W/Z$ cross section, while the s-channel (essentially a Drell-Yan process) is insensitive to the $b$-quark PDF and can therefore act as a control, and it is correlated with the $W/Z$ cross section, like the \ttbar\ cross section~\cite{guffanti-rojo}.
 Moreover, the charge asymmetry in t-channel production as a function of the top/antitop rapidity will provide a very powerful input for constraining PDFs, similar to the $W$ production case.

Different beyond-SM models predict different effects in the different production channels~\cite{cpyuan}.
 A few examples follow, reasoning in a bottom-up approach (i.e., starting from the possible observations to infer which model would be favored):
\begin{itemize}
\item a deficit in all three channels would naturally lead to suspect $|V_{tb}|<1$ (and the existence of new quarks)~\cite{vtb}; this can be verified by precisely measuring $R_b\equiv\frac{BR(t\to Wb)}{BR(t\to Wq)}$ in \ttbar\ events (section~\ref{sec:couplings});
\item an excess in the s-channel, not confirmed in the other two, would induce to suspect a charged resonance (e.g., a right-handed $W^\prime$, whose coupling to light fermions would be suppressed by helicity conservation and would therefore be visible in the $t\bar b$ final state and not in electronic or muonic decays), which would be confirmed by a peak or a peak-dip structure in the $M_{t\bar b}$ spectrum;
\item an excess in the t-channel, not confirmed in the other two, could be due to new interactions causing flavour-changing neutral currents (FCNC): even with tiny $ut\gamma$ and $utZ$ couplings, the very large up-quark density at high $x$ in the proton would allow a visible signal to show up; important checks would be the differential measurements of $d\sigma^{\rm t-channel}_{t/\bar t}/dy$ and of the single top polarization (section~\ref{sec:spin});
\item an excess in the t-channel and in the $tW$ channel, with no deviation from the SM in the polarization of the top quark, and a deficit in the s-channel could be due to large $|V_{td}|$ or $|V_{ts}|$ (non-unitarity of the CKM matrix), to be checked by measuring $d\sigma^{\rm t-channel}_{t/\bar t}/dy$ and $R_b$.
\end{itemize}

\section{Top quark polarization and spin correlations}
\label{sec:spin}

A crucial property of the top quark is its lifetime which is shorter than the QCD de-coherence timescale, causing its decay products to retain memory of its helicity.
 This provides additional powerful tools in the search for new physics, both in \ttbar\ and single top studies:
 single top production in t- and s-channel yields 100\% polarized top quarks if only electroweak interactions are involved, while different new physics models yield different polarizations; pair-produced tops are unpolarized but theirs spins are correlated, and the spin of a neutral resonance decaying in \ttbar\ can be directly inferred by the spin correlation, which in the SM depends uniquely on PDFs (fraction of $gg/qq/qg$ initial states) and provides a handle to constrain them under the assumption of no new physics.

The ATLAS collaboration presented a first study of \ttbar\ spin correlations~\cite{spincor_atlas}, showing a better than 40\% precision on the correlation coefficient with 0.7 \fb\ in dileptonic events by fitting the difference in azimuthal angle between the two charged leptons.
  Extrapolating from older simulation studies with full event reconstruction, $\pm$10\% seems realistic with the expected 2012 dataset in both the dilepton and single lepton channels~\cite{tdr_cms,tdr_atlas}.

Single top polarization in the t-channel is already manifest in the slope of the $\cos\theta^*_{lj}$ distribution, defined as the angle between the charged lepton and the light jet in the reconstructed top quark rest frame. The SM expectation $\cos\theta^*_{lj}\approx 1$~\cite{mahlon} has even been exploited in one of the CMS cross section analyses~\cite{st_cms} to better identify the SM signal; the next generation analyses, with more data, will have the possibility of a precision measurement of $d\sigma/d\cos\theta^*_{lj}$, sensitive to new physics (section~\ref{sec:st}).

\section{Top quark couplings}
\label{sec:couplings}

Information on the $tbW$ vertex comes from several complementary inputs.
 The cross section of all single top channels (section~\ref{sec:st}) measures ``directly'' $|V_{tb}|$ under the assumption that $R_b = 1$ and the coupling is purely left-handed ($f_L=1$), while $R_b$ measures $|V_{tb}|$ under the assumption that the CKM $3\times 3$ matrix is unitary and that  $f_L=1$; ATLAS and CMS will soon challenge the most precise measurement from D0, $R_b=0.90\pm 0.04$~\cite{br_d0}. The combination of $R_b$ with all single top cross sections would give a less model-dependent handle on $|V_{tb}|$~\cite{vtb} although still with the $f_L=1$ assumption.

A handle on the anomalous components of the $tbW$ coupling comes from the polarization of the $W$ boson from top quark decays in \ttbar\ events, already measured by ATLAS with 0.7 \fb~\cite{wasym_atlas}. The combination of this information with the single top cross section is sufficient to set constraints in the $(f_L,f_R,g_L,g_R)$ hyper-plane under the assumption $|V_{tb}|\approx 1$~\cite{aguilar2}.
 An improvement of more than a factor 2-3 in the precision of either of the two inputs at the LHC is needed to improve over the limits from Tevatron, and is realistic with less than 10 \fb.

Another goal for 2012 is the improvement of the existing limits on the forbidden FCNC couplings  $tqV^0$ with $q=u,c$ and $V^0=\gamma,Z,g$.
 First studies have been performed by ATLAS with 35 \pb~\cite{fcnc_atlas} with two strategies: limits on $tqZ$ and $tq\gamma$ couplings have been set by searching for $t\bar t\to WbZq$ and $t\bar t\to Wb\gamma q$ events (with $W\to l\nu$ decay), and on $tqg$ in the $qg\to t\to l\nu b$ process.
 With 10 \fb\ the LHC experiments have the potential to improve the limits set at previous accelerators (LEP, HERA, Tevatron). Complementary constraints on $tqZ$ and $tq\gamma$ couplings would come from single top measurements, sections~\ref{sec:st} and~\ref{sec:spin}.

\section{Top quark mass}
\label{sec:mass}

It will be very challenging for the LHC experiments to improve the 0.5\% precision on $m_t$ achieved at the Tevatron~\cite{mass_tevatron}. The traditional direct reconstruction methods are systematics-limited, therefore a tenfold increase in statistics doesn't help {\it per se}.

 A crucial direction for improvement is the study of additional methods uncorrelated with the traditional analyses.
 For example, CDF pioneered methods for mass extraction based on the charged lepton spectrum~\cite{cdf_pt} or the $b$-hadron decay length~\cite{hill,cdf_lxy} or both~\cite{cdf_pt_lxy}. Both observables have the advantage of negligible dependence on the jet energy scale, which is the limiting factor in direct reconstruction methods. The lepton-spectrum method offers the further advantage, with the $W$ boson being emitted before QCD enters the picture, that the mass that it measures is conceptually closer to the pole mass.
 While these studies were statistics limited at the Tevatron and were therefore not used in the combination, the LHC experiments have the double advantage of a much larger dataset and of a better purity. Despite additional complications coming from the significant boost of the top quarks, a precision of $\Delta m_t < 2$~GeV seems realistic with 10 \fb\ at 7 TeV.

An original technique proposed in CMS simulation studies~\cite{tdr_cms} and never applied so far in data exploits the $t\to Wb\to l+J/\psi +X$ decay chain ($5.5\times 10^{-4}$ branching ratio, yielding around 900 events in 10 \fb\ at 7 TeV) and the correlation between $m_t$ and $M(l,J/\psi)$. The dependence on the jet energy scale is negligible, $b$ tagging is not needed (the presence of a $J/\psi$ is sufficient to reject most light jets), and the systematic uncertainty ($\approx 1.5$~GeV) is dominated by the modeling uncertainties, especially fragmentation.

\section{Prospects for a 2012 run at higher energy}

At the time of writing a possible increase to 8 TeV during the 2012 run is under discussion.
 Running at higher energies is in general less crucial for SM precision measurements than for searches of new physics, but testing the dependence of the cross sections versus \sqrts\ will be an important test of the production models.
 The signal-to-background ratio improves from 7 to 8~TeV: \ttbar\ (figure~\ref{fig:sqrts}, left) and $tW$ cross section would increase by 50\% and t-channel single top by 30\%  (figure~\ref{fig:sqrts}, right), against a 20\% increase for inclusive $W$ production and 25\% for the $WQ(\bar Q)$ process (with $Q=b,c$).
 In addition, analyses can take advantage from the different dependences of these processes on energy, e.g., by simultaneous fits to 7 and 8~TeV data with $\sigma_{bkg}({\rm 8~TeV})/\sigma_{bkg}({\rm 7~TeV})$ ratios constrained to SM expectations.

\section{Conclusions}

With an accumulated statistics of more than 10 \fb\ (either all at 7 TeV, or shared between 7 and 8 TeV) the LHC will be even more established as ``the top factory'', and top physics itself will be even more established as a powerful tool for direct and indirect new physics searches, with strong complementarity between the different production modes.
 Although most of the precision measurements start already to be systematics-limited, larger datasets would help in many ways: tighter selections can be applied, high-statistics sidebands can better constrain the main systematic uncertainties, and orthogonal analyses with low rate may become competitive in the combination with standard methods. In addition, the extensive validation of the background models and the better understanding of most experimental and modeling systematics will allow more dedicated analysis techniques. 
 For all these reasons, plus the confidence in the ingenuity of the experimental community, we can confidently look forward to an unprecedentedly rich top physics program to be presented at TOP2012.

\acknowledgments
The author wishes to thank the TOP2011 organizers and participants for the wonderful atmosphere, Roberto Tenchini, Roberto Chierici, Jeannine Wagner-Kuhr, Joe Incandela and Ford Garberson for many useful inputs, and the Estonian Academy of Science for supporting this work with the Mobilitas Top Researcher Grant MTT59.


\begin{thebibliography}{0}
\bibitem{atlas} \BY{ATLAS Coll.} 
 \IN{JINST}{3}{2008}{S08003} 
\bibitem{cms} \BY{CMS Coll.} 
 \IN{JINST}{3}{2008}{S08004} 
\bibitem{uwer} \BY{Uwer~P.} 
 these proceedings
\bibitem{xsec_ttbar_atlas} \BY{ATLAS Coll.} 
 ATLAS-CONF-2011-121
\bibitem{xsec_ttbar_cms} \BY{CMS Coll.} 
 CMS PAS TOP-11-024
\bibitem{mtt_atlas1} \BY{ATLAS Coll.} 
 ATLAS-CONF-2011-123
\bibitem{mtt_atlas2} \BY{ATLAS Coll.} 
 ATLAS-CONF-2011-087
\bibitem{mtt_cms1} \BY{CMS Coll.} 
 CMS PAS EXO-11-055
\bibitem{mtt_cms2}\BY{CMS Coll.} 
 CMS PAS EXO-11-006
\bibitem{frederix2009} \BY{Frederix~R. \atque Maltoni~F.} 
 \IN{JHEP}{0901}{2009}{047}
\bibitem{boosted1} \BY{Bazterra~V.} 
 these proceedings
\bibitem{boosted2} \BY{Spannowsky~M.} 
 these proceedings
\bibitem{afb_cdf1} \BY{CDF Coll.}
 \IN{Phys. Rev. D}{83}{2011}{112003}
\bibitem{afb_cdf2} \BY{CDF Coll.}
 CDF Note 10436
\bibitem{afb_d0} \BY{D0 Coll.}
 arXiv:1107.4995 [hep-ex]
\bibitem{aguilar} \BY{Aguilar-Saavedra~J.-A.}
 these proceedings
\bibitem{asym_cms}\BY{CMS Coll.}
 CMS PAS TOP-11-014
\bibitem{asym_atlas} \BY{ATLAS Coll.} 
 ATLAS-CONF-2011-106
\bibitem{aguilar-juste-rubbo}\BY{Aguilar-Saavedra~J.-A. \atque Juste~A. \atque Rubbo~F.}
 arXiv:1109.3710 [hep-ph] 
\bibitem{ttgamma_atlas} \BY{ATLAS Coll.} 
 ATLAS-CONF-2011-153
\bibitem{tdr_cms}\BY{CMS Coll.}
 \IN{J. Phys. G: Nucl. Part. Phys.}{34}{2007}{995}
\bibitem{tdr_atlas}\BY{ATLAS Coll.}
 CERN-OPEN-2008-020
\bibitem{ttH}\BY{Allwood-Spiers~S.}
 these proceedings
\bibitem{ttNj_atlas}\BY{ATLAS Coll.}
 ATLAS-CONF-2011-142
\bibitem{gerber}\BY{Gerber~C.}
 these proceedings
\bibitem{st_cms}\BY{CMS Coll.}
 \IN{ Phys. Rev. Lett.}{107}{2011}{091802}
\bibitem{st_atlas}\BY{ATLAS Coll.}
 ATLAS-CONF-2011-101
\bibitem{motylinski}\BY{Motylinski~P.}
 these proceedings
\bibitem{tw_cms}\BY{CMS Coll.}
 CMS PAS TOP-11-022
\bibitem{tw_atlas}\BY{ATLAS Coll.}
 ATLAS-CONF-2011-104
\bibitem{ss_atlas}\BY{ATLAS Coll.}
 ATLAS-CONF-2011-118
\bibitem{guffanti-rojo} \BY{Guffanti~A. \atque Rojo~J.}
 arXiv:1008.4671
\bibitem{cpyuan} \BY{Tait~T. \atque Yuan~C.-P.}
 \IN{Phys. Rev. D}{63}{2001}{014018}
\bibitem{vtb}\BY{Alwall~J. \atque Frederix~R. \atque G\'erard~J.-M. \atque Giammanco~A. \atque Herquet~M. \atque Kalinin~S. \atque Kou~E. \atque Lema\^itre~V. \atque Maltoni~F.} 
 \IN{ Eur. Phys. J. C}{49}{2007}{791-801}
\bibitem{spincor_atlas} \BY{ATLAS Coll.} 
 ATLAS-CONF-2011-117
\bibitem{mahlon}\BY{Mahlon~G. \atque Parke~S.} 
 \IN{Phys. Lett. B}{476}{2000}{323}
\bibitem{br_d0} \BY{D0 Coll.} 
 \IN{Phys. Rev. Lett.}{107}{2011}{121802}
\bibitem{aguilar2} \BY{Aguilar-Saavedra~J.-A., Castro~N.~F., Onofre~A.} \IN{Phys. Rev. D}{83}{2011}{117301}
\bibitem{wasym_atlas} \BY{ATLAS Coll.} ATLAS-CONF-2011-037
\bibitem{fcnc_atlas} \BY{ATLAS Coll.} ATLAS-CONF-2011-061
\bibitem{mass_tevatron}\BY{Tevatron Electroweak Working Group} arXiv:1107.5255
\bibitem{cdf_pt}\BY{CDF Coll.} \IN{Conference Note}{9881}{2009}{}
\bibitem{hill}\BY{Hill~C.~S. \atque Incandela~J. \atque Lamb~J.~M.} \IN{Phys. Rev. D}{71}{2005}{054029} 
\bibitem{cdf_lxy}\BY{CDF Coll.} \IN{Phys. Rev. D}{75}{2007}{071102} 
\bibitem{cdf_pt_lxy} \BY{CDF Coll.} \IN{Phys. Rev. D}{81}{2010}{032002} 
\end{thebibliography}
\end{document}